\documentclass[useAMS,usenatbib,usegraphicx,amsmath]{mn2e}

\title[Planetesimal accretion around binaries]
{Relative velocities among accreting planetesimals 
in binary systems: the circumbinary case}

\author[H. Scholl, F. Marzari, P. Th\'ebault]
{H. Scholl$^{1}$\thanks{E-mail:Hans.Scholl@oca.eu},
F. Marzari$^{2}$, P. Th\'ebault$^{3}$\\
$^{1}$Laboratoire Cassiop\'ee, CNRS, Observatoire de la C\^ote d'Azur, B.P. 4229,
F-06304 Nice, France\\
$^{2}$Department of Physics, University of Padova, Via Marzolo 8,
35131 Padova, Italy\\
$^{3}$Stockholm Observatory, Albanova Universitetcentrum,
SE-10691 Stockholm, Sweden, and\\ 
Observatoire de Paris, Section de Meudon, F-92195 Meudon Principal Cedex,
France}

\begin{document}

\date{Accepted .... Received .....; in original form ...}

%\pagerange{\pageref{firstpage}--\pageref{lastpage}} \pubyear{...}

\maketitle
\label{firstpage}

\begin{abstract}
\noindent

We numerically investigate the possibility of planetesimal accretion
in circumbinary disks, under the coupled influence of both stars' secular
perturbations and friction due to the gaseous component of the
protoplanetary disk. We focus on one crucial parameter: the
distribution of encounter velocities between planetesimals in
the 0.5 to 100km size range. An extended range
of binary systems with differing orbital parameters is explored.
The resulting encounter velocities are compared to the threshold
velocities below which the net outcome of a collision is accumulation into
a larger body instead of mass erosion.
For each binary configuration, we derive the critical radial
distance from the binary barycenter beyond which planetesimal accretion
is possible. This critical radial distance is
smallest for equal-mass binaries on almost circular orbits. It
shifts to larger values for increasing eccentricities and decreasing mass
ratio. The importance of the planetesimals' orbital alignments of
planetesimals due to gas drag effects is discussed.

\end{abstract}


\section{Introduction}

At present more than 40 planets have been found in binary star systems, which
corresponds to
about 20 \% of the whole sample of known extrasolar planets \citep{desi07}.
All of them are on so--called S--type or circumprimary orbits 
that encircle one component of the double star
system. Planets may also revolve on dynamically stable orbits
about both binary stars on 
so--called P--type or circumbinary orbits (e.g. Holman and Wiegert 1999). 
At present, there is only one
planet assumed to revolve on a circumbinary orbit. The primary, HD 202206, is
a metal rich main sequence star. Its companion, HD 202206b, is assumed to be
a low--mass brown--dwarf with an estimated mass of 17.5 $M_{Jup}$ according
to radial velocity measurements \citep{udry02,corr05}.
This mass is beyond the widely accepted brown-dwarf limit of 10 Jupiter masses.
The third body, HD 202206c, is assumed to be a Jupiter--like planet 
revolving about the binary.
Adopting a mass of 2.41 $M_{Jup}$ for HD 202206c, 
Correia et al. (2005) show in a dynamical analysis that 
HD 202206c and HD 202206b might be in a 5/1 mean motion resonance.  
This resonant configuration
is possibly 
a consequence of an inward migration of HD 202206c 
due to the forces exerted by a
viscous circumbinary disk (Nelson 2003). 

The signature of 
circumbinary planets in radial velocity measurements is in most cases difficult to be 
spotted because of the short--term large--amplitude  velocity fluctuations 
on the primary induced by the companion star. 
However, planets might be very frequent
among the population of close star couples. 
Mid--infrared emissions have in fact revealed 
substantial circumbinary 
material around PMS close binaries like DQ Tau, UZ Tau, GW Ori
(Mathieu et al. 2000) and AK Sco \citep{jema},
and more detached binaries like 
GG Tau \citep{du94}, UY Aur \citep{close}.
These disks are significantly more massive than the minimum--mass solar nebula
which suggests abundant planetary formation.
However, both concurrent mechanisms for extrasolar
planet formation, namely the standard solid--core model forming planetary
embryos from runaway and oligarchic accretion of planetesimals
\citep[e.g][]{gre78,wet89,kok00} possibly followed by 
gas infall onto the core for the formation of giant planets (Pollack et al. 1996; 
Bodenheimer and Lin 2002), and the alternative model of local gravitational
collapse of disk material \citep{bos1} might be affected by the secular
perturbations of the binary. 

We will focus here specifically on the ``standard'' solid--core scenario.
in particular on the stage of planetary embryo accretion. This stage is just
before the one investigated by Quintana and Lissauer (2006) who determined
regions around close binaries where planets can form by accumulation of embryos.
Within the frame of this model, 
the initial stage of planetesimal accretion 
is the runaway growth of isolated planetary embryos.
This stage is very fast and efficient,
provided that the encounter velocities among planetesimal are
low, i.e. much smaller than the escape velocities of the growing
embryos \citep{lis93}.
As a consequence, runaway growth is particularly sensitive to 
any external perturbation able to stir up  relative velocities in 
the planetesimal swarm. In the specific case of a circumbinary disk,
relative velocities between planetesimals can be expected to be higher
as compared to the single star case. In the vicinity of the binary,
velocities could be increased to values for which runaway accretion is
no longer possible, or even to higher values for which
collisions may result in mass removal via cratering or
fragmentation rather than accretion. 
The threshold velocity for the stop of runaway is of the order
of the escape velocities $v_{esc(R)}$ of the growing bodies while
the threshold velocity $v_{ero}$ for the limit between accreting
and eroding impacts is less straightforward to estimate and
depends on the respective sizes and physical properties of the two colliding
planetesimals.
In any case, the distribution of encounter velocities 
is the key parameter controlling the fate of a collisionally interacting
swarm of planetesimals.
This is the issue which we address here, by numerically estimating
the distribution of $<\Delta v>$
in a disk of planetesimals surrounding a binary system in order to derive 
assertive conclusions on the possibility of planet formation. 

Only taking into account the gravitational potential of both stars
might lead to incorrect estimates of relative velocities, in particular 
close to the star couple. Indeed, the planetesimal accretion phase
is believed to take place while a large fraction of the primordial
gas disk is still present. 
Gas drag is known to circularize orbits \citep{ada76} and thus
reduce relative velocities in dynamically hot systems \citep{mascho00}.
Besides circularization,
gas drag has the additional effect of aligning
orbital apsides, an effect which is also called orbital synchronization
\citep{mascho97,mascho00}.
The consequences of this synchronization on relative velocities
depend on planetesimal sizes. Bodies
of the same size all have their periapses aligned in the same direction. Smaller
bodies have a much smaller dispersion of periapse directions than larger bodies.
Furthermore, the mean direction of the periastron alignment is also
different for objects of different sizes \citep{the06}.
As a result, while
relative velocities between small planetesimals of the same size may drop to
almost zero, relative velocities among bigger bodies are more significant because
of a larger periapse direction dispersion. The highest relative velocities 
can be expected between
objects of {\it different} sizes due to the size--dependency of
the periastron alignment.
The relative velocity distribution depends, therefore, not only 
on the radial distance to the barycenter of the binary but also on the sizes of
colliding planetesimals and on the gas density. 

In a pioneering attempt 
to model planetary accretion in circumbinary disks, Moriwaki and 
Nakagawa (2004) used a  purely gravitational model without gas drag.
They found very stringent limits on the minimum distance 
from the binary beyond which planetesimal accumulation is possible.
However, 
as pointed out above, neglecting the velocity damping effect 
and orbit alignment due
to gas friction may yield misleading results for the 
efficiency of planetesimal accumulation, in particular in the vicinity of 
the binary. 
In the present work, we redress this issue by performing 
numerical simulations where the effects
of gas drag are taken into account, although with unavoidable
simplifying assumptions. Moreover, we compute the average
encounter velocities between planetesimals from
their trajectories and not from the values of forced eccentricities
as \citet{mori} who used the 
relation $\Delta v \simeq e.v_{kep}$, where $v_{kep}$ denotes Keplerian
velocity.
This often used relation holds only under the condition that all 
orbits have fully randomized Keplerian angles. However, in circumbinary
disks with gas, orbital phasing of periastra directions is  of high importance 
because of 
the combined effects of secular perturbations and gas friction. 
Another limitation of the $\Delta v \simeq e.v_{kep}$ relation is that it
works $locally$ and it does not take into account the possibility of  
large radial excursions of bodies with high eccentricities \citep[for more
on this subject, see for instance the discussion in][]{thedo03}.
Both these issues suggest that the relation used by Moriwaki and Nakagawa (2004)
may not be appropriate in presence of large orbital eccentricities 
and possibly of gas drag. 
Our numerical approach, which takes into account all close particle encounters
representing collisions,
has the advantage to handle automatically 
orbital phasing (and dephasing) as well as radial mixing effects.
This allows us to derive more reliable values for 
the limiting distance from the barycenter of the binary
where planet formation is possible
for different parameters of the binary. Not surprisingly,
our results significantly differ from those of Moriwaki and Nakagawa (2004)
in particular close to the binary where planetary accretion is more likely to
form planets due to the higher density of material.

Due to obvious numerical constraints, mutual 
gravitational interactions among planetesimals, like for instance dynamical friction,
are not included even if, in principle, they might be taken into account in
N-Body simulations. 
However, the speed of current
computers is still far too low to allow a systematic investigation of binary systems
surrounded by a self-gravitating planetesimal disk.
These simulations are still restricted to
disks with massless planetesimals. Their size distribution, of course, does not
evolve with time. To do so, particle--in--a--box computations,
which can follow the 
evolution of a large number of massive bodies, are required,
but the price to pay is a significant loss of accuracy in the computations of the 
orbital evolution of the planetesimals that, in the present scenario, is
crucial. As a consequence, we decided to adopt a 
deterministic numerical approach, which, although
neglecting important physical effects, gives a reliable
estimate of encounter velocity distributions and thus enable to
derive ``maps'' of the 
disk regions with relative velocities low enough to favor
planetesimal accretion and planet formation.

Our numerical model is described in detail in Section 2.
Section 3 is devoted to the computations performed
in the gas-free case, a scenario similar to that of \citet{mori}.
In Section 4 we use the results obtained from models including gas drag in
order to map the regions where we expect planet formation by accretion.
Section 5 is dedicated to the dicussion of our results.

\section{The numerical model}

\subsection{Main characterstics}

Our deterministic code is an upgraded version from the one used in several
earlier studies \citep{mascho98,mascho00}, adapted
to the specificities of the circumbinary configuration.
We consider a system of $N\sim 10000$ planetesimals, treated as non--gravitationally
interacting test particles, submitted to the gravitational potentials
of both stars and to a gas drag force following \citet{ada76} and
described in the following section.
Planetesimal orbits are integrated using a fourth-order
Runge-Kutta scheme with fixed step size.
We like to point out that for the circumprimary
problem \citep{the06} we applied a different numerical scheme which is
less convenient for the circumbinary case.

The binary stars move on an elliptic orbit of eccentricity $e_b$
and semi--major axis $a_b$. For sake of simplicity and to limit
the number of free parameters to a manageable value, we assume
for all runs $a_b=1\,$AU. This is the standard value
also taken by \citet{mori} in their numerical explorations.
We are thus here implicitly focusing on planetesimal accretion
in the regions typically beyond 3--4\,AU from the
binary's centre of mass. This region is sufficiently away from the critical
instability region found by Holman and Wiegert (1999) to guarantee orbital stability
for almost planar and circular motion.
Both stars' total mass is fixed and equal to 1 solar mass. 
The primary parameters of our model are thus the mass ratio and eccentricity
of the binary star system.
We use five different mass ratios $q$ for the binary, 
$q$ = 0.5, 0.4, 0.3, 0.2 and 0.1. The stellar masses $m_1$ of the primary
and $m_2$ of the companion are then $m_1 = 1 - q$ and $m_2 =  q$, 
in solar mass units. 
For each mass ratio, models
with six different binary eccentricities $e_b$ are simulated with $e_b$ = 0.0, 0.1,
0.2, 0.3, 0.4 and 0.5. This gives a total of 30 models. Each model is simulated
with and without gas drag.
In models taking into account gas drag, we consider
planetesimal  of physical radii ranging from 0.5km to 100km.
The initial eccentricities and orbital inclinations are chosen with a uniform
random distribution between 0 and $10^{-5}$.
The initial particle semi--major axis are randomly distributed
between $a_{min}=4\,$AU and $a_{max}=12\,$AU. The value for $a_{min}$
being approximately equal to the inner truncation 
of a circumbinary disk by the tidal influence of the
binary \citep{arty94, gukl}. Whenever the semimajor axis 
of a planetesimal decreases below the inner limit
because of drag friction, the body is removed. 

The typical timescale for one run is $\sim$100 000 binary revolutions,
which is approximately the time necessary to reach a steady state, with
fully developed secular perturbations and the gas drag induced orbital
phasing.
After such a steady state is reached, we determine relative velocities
among the planetesimal population.
For this purpose, we use the positions
and velocities of all planetesimals at a fixed time. For each planetesimal,
we search its nearest neighbouring planetesimal and compute the relative velocity
corrected for Kepler shear due the different radial distances. Relative velocities
 are sampled in bins in radial direction. A bin size of 0.5 AU is used. The median
value for each bin is then taken as the representative value for the collisional
or impact velocity of planetesimals
at the corresponding radial distance. 
Since, as pointed out, the orbit alignment is size dependent, we compute the median
relative velocities $\Delta v_{(R1,R2)}$ between planetesimal populations for each
pairs of radii $R_1$, $R_2$ in each bin.
We typically use 20 to 30 close encounters in a bin to determine the 
corresponding median relative velocity.

\subsection{The gas drag force}

The drag force is modelled in laminar gas approximation by

\begin{equation}
\label{eq:dra}
\ddot{\bf r}  = - K v_{rel} {\bf v_{rel}},
\end{equation}
where $K$ is the drag parameter given by \citep{karo} as
\begin{equation}
\label{eq:dragpa}
K=3 \frac{\rho_g C_d} {8 \rho_{pl} s}
\end{equation}
where $s$ is the planetesimal radius, 
$\rho_{pl}$ its mass density, $\rho_g$ the gas density of the 
protoplanetary disk and $C_d$ a dimensionless drag coefficient
related to the shape of the body ($\sim 0.4$ for spherical bodies).
Exploring the gas drag density profile as a free parameter would
be too CPU time consuming and we shall restrict ourselves to one
$\rho_{\rm g}$ distribution. We
assume the standard Minimum Mass Solar Nebula (MMSN) of \citet{haya81},
with $\rho_{\rm g}=\rho_{\rm g0}(a/1AU)^{-2.75}$ and
$\rho_{\rm g0}=1.4\times10^{-9}$g.cm$^{-3}$. 
We take a typical value $\rho_{\rm pl}=3\,$g.cm$^{-3}$.

We are here implicitly assuming that the gas disk is axisymmetric 
and pressure supported.
We are aware that this axisymmetric prescription does probably not
represent the real physical structure of the circumbinary gaseous
component of the disk.
This is a complex issue, which has only been addressed in
a handful of elaborated numerical investigations
\citep[e.g][]{arty94,gukl,nel03,pich05}. All these studies clearly show the
inner tidal truncation of the disk, but are less detailed on 
the structure within the disk, which is particularly 
complex to be precisely computed, especially in the
eccentric binary case. The only clear result is
the onset of complex time--evolving spiral patterns extending from the 
inner edge of the disk towards the outer regions.
At present it appears a very demanding task to perform a fully coupled
numerical treatment of both the gas disk and an embedded planetesimal swarm.
For this reason, we prefer in a first step a simplified approach 
where the gas drag force is just described by Equ.\ref{eq:dra}.
The onset of density waves in the disk induces 
radial and azimuthal variations both
of the disk density and of the gas velocity
where an additional radial component adds up to 
the Keplerian tangential component. These effects might
alter the strength of the drag acceleration          
when a planetesimal enters and exits the higher or lower density region.
However, it is reasonable to assume that, on the average, this effect
would not strongly modify accelerations obtained by Equ.\ref{eq:dra}
and thus the dynamical evolution of planetesimals,
at least for planetesimals in the km--size range.
In particular, we expect the occurrence of 
the periastron alignment.

As already mentioned, we start all planetesimals on almost circular orbits.
This choice is from a dynamical point a view equivalent to assuming
that, at $t=0$, the proper eccentricity is almost equal to the forced one $e_f$.
This is a somewhat artificial and not a necessarily realistic starting configuration
motivated by the fact that, at present, we have no generally accepted clues
either on how planetesimals form within a disk with spiral waves 
nor on the initial orbital elements of the planetesimals as they
detach from the gas. 
We integrate the model until a steady
state is reached where periapses are aligned and the interaction with the gas becomes
only a minor perturbation of the Keplerian orbit. This procedure is valid for 
small planetesimals, since they reach the same periapse alignment
independently of their starting values with the same distribution of 
proper eccentrcities.
Larger planetesimals, which are much less affected by gas drag,
have their proper eccentricities increasing
during their growth by collisions and
close encounters with other large planetesimals. As a consequence,
they too should lose memory of their initial orbits.
As a consequence, we think that
our initial choice of eccentricities does not introduce too
significant effects on the calculation of relative impact velocities
once the steady state is reached.

\section{Impact velocities in the gas free case}

\begin{figure}
\includegraphics[width=8.5cm]{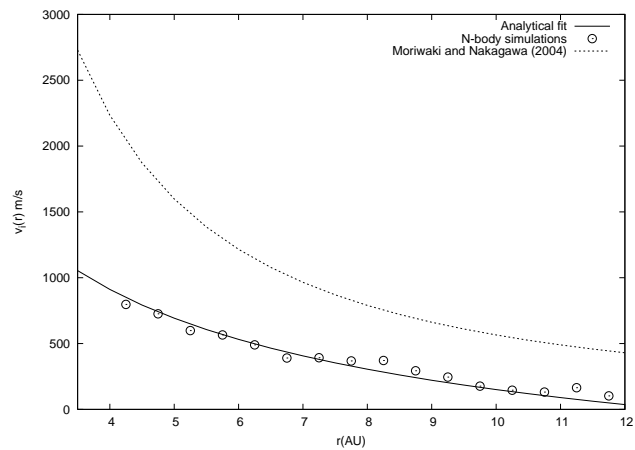}
\caption{Relative impact velocities between planetesimals computed 
with our equation~\ref{eq:fit} (continuous line) compared 
to the predictions of \citep{mori} (dashed line). The empty 
circles show the numerical results of the N-body integrations.
The parameters of
the binary are $q=0.2$ and $e_b = 0.4$.}
\label{f1}
\end{figure}

In the gas free models we use 25000 particles with semimajor axes ranging
from 3 to 50 AU. However, we focus here essentially on the region within
12AU, since, as it clearly appears in Fig.\ref{f1}, 
average encounter velocities drop down to almost zero beyond this limit.
For the $r<12\,AU$ region, the $\langle \Delta v \rangle$ we obtain might
be fitted, for the range of stellar masses are orbital eccentricities
explored, by the empirical formula 

\begin{eqnarray}
\label{eq:fit}
\langle \Delta v \rangle_{(r)} & = & (-384.2+2067.7 q-447522 e_B \nonumber \\
	&  &   -3864.9 q^2+9462.4 q e_B) \nonumber \\
	&  & +(1024.8-5783.4   + 16431 e_B \nonumber \\
	&  & +11251 q^2-34396 q e_B)\frac {1} {\sqrt{r}}
\end{eqnarray}
which is a second-order polynomial in
$q$ and $e_b$ multiplied by an additional term 
equal to $\sqrt(r)$ accounting for the radial dependence of
$\langle \Delta v \rangle$ on the local Keplerian velocity. 

Since we neglect the gas drag force, the radius of 
the planetesimals is not a parameter of the problem.
In Fig.\ref{f1} we also compare our results to the
$\langle \Delta v \rangle$ computed with the \citet{mori} analytical 
approximation  for the specific case 
with $q = 0.2$ and $e_b = 0.4$. For reference, we also plot
the numerical data obtained from the N-body integration.  
It can be clearly seen that our numerically derived values
are always significantly smaller than the \citet{mori} estimates.
This shows the weakness of the 
approximate relation $\Delta v \simeq e v_{kep}$ for the
present case, where orbital phasing and radial excursions 
of the planetesimals play a crucial role in determining
the average impact velocity. As an example, while our simulations 
give low relative velocities, of the order of 50 m/s at about 11
AU from the barycenter, the analytical approximation of \citet{mori}
predicts such low velocities only beyond 50 AU.
Our numerical results give much lower values for impact velocities
which allow planetary formation to occur much closer to 
the barycenter of the binary even in absence of gas friction. This 
is a first indication that planets on circumbinary orbits may be abundant
due to the low relative velocities in the regions near the binary with 
higher surface density. In the following section, we show how
this promising result is affected by gas drag.

\section{Impact velocities in the presence of gas}
\label{accgas}

We now turn to the more realistic case including gas drag.
As previously mentioned, it is well known that
the combination of the drag force
and perturbations by the gravity field of the binary causes 
a strong paeriapse phasing 
of the orbits of the planetesimals. The planetesimals' physical
sizes are now a crucial parameter. Bodies of similar sizes have their
periapses aligned towards approximately the same direction, which 
leads to very low impact velocities. Different size planetesimals 
are aligned towards different directions, increasing the 
non--tangential component of the impact velocities. 

To compute the relative velocity between any possible pair
of different sized planetesimals in the swarm is far 
beyond the present computing capabilities. Therefore, we 
derive the mean relative velocities $\Delta v$ for two 
separate cases assumed to be representative for the 
accreting capability of a swarm:
\begin{itemize}

\item  $\Delta v_{R1, R1}$ between equal-size bodies 
of radius $R_1$. In this case the periapse alignment is 
stronger. 

\item  $\Delta v_{R1, R2}$ for a target with radius 
$R_1$ and a projectile with radius $R_2 = R_1/2$. This size ratio
has been chosen since it is the configuration for which,
in the large majority of cases explored in our circumprimary
study \citep{the06}, the kinetic energy delivered by the impactor is maximal.
For $R_1/2 <R_2 <R_1$ the relative velocity
$\Delta v_{R1, R2}$ decreases and approaches the low velocity 
value of equal--size impacting bodies. 
For $R_2<R_1/2$, the increase
in $\Delta v_{R_1, R_2}$ is too slow to compensate for the 
diminishing impactor mass. 
\end{itemize}

\begin{figure*}
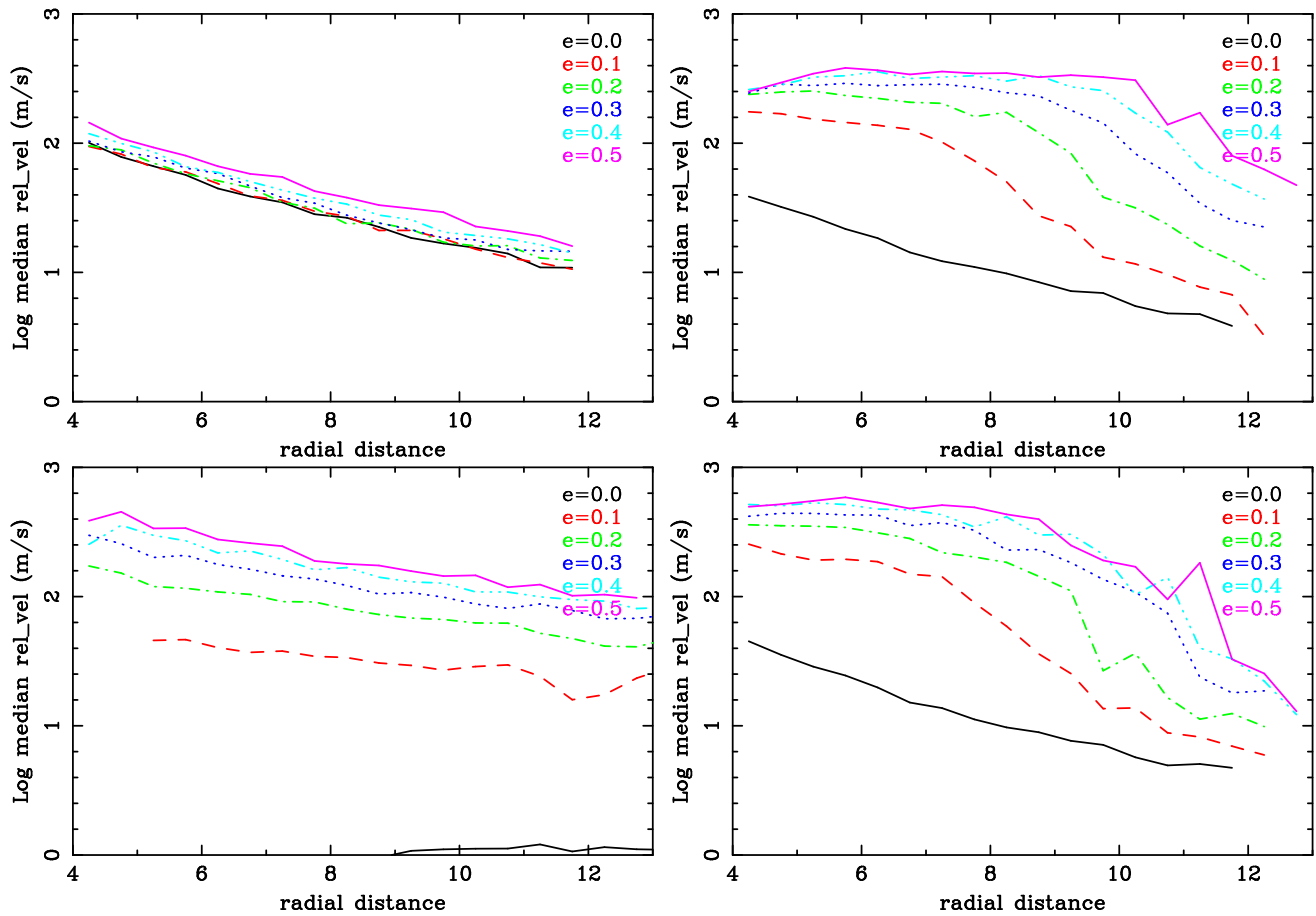

\makebox[\textwidth]{
\includegraphics[width=6cm,angle=270]{snap1.ps}
\hfil
\includegraphics[width=6cm,angle=270]{snap2.ps}
}
\makebox[\textwidth]{
\includegraphics[width=6cm,angle=270]{snap3.ps}
\hfil
\includegraphics[width=6cm,angle=270]{snap4.ps}
}
%\makebox[\textwidth]{
%\includegraphics[width=\columnwidth]{snap5.ps}
%\hfil
%\includegraphics[width=\columnwidth]{snap11.ps}
%}
\caption[]{Figure showing the relative velocity between planetesimals
for different values of the binary eccentricity $e_b$, mass ratio
$q$ and planetesimal sizes $R_1$ and $R_2$. The plots are organized 
as follows: top left $q=0.5$ (equal--mass stars case),
$R_1 = 25$ km and $R_2=50$km; 
top right $q=0.1$, $R_1 = 25$ km and $R2=50$km; 
bottom left $q=0.1$, $R_1 = 0.25$ km and $R_2=0.5$km;
bottom right $q=0.1$, $R_1 = 100$ km and $R_2=50$km.
}
\label{snapsh}
\end{figure*}

For each of the 30 different ($q$,$e_b$) binary configurations
considered, we perform 6 simulations exploring different 
target--impactor pairs
($R_1$,$1/2R_1$) in the $0.5$km$<R_1<100$km
range (leading to a total of 180 runs). We compute for each pair 
the average impact velocity $<\Delta v>_{R1,R2}$. 
In Fig.\ref{snapsh} we show some examples of the relative 
velocity between different size planetesimal swarms. 
A general feature is that
binaries with higher eccentricities always lead to 
higher $\langle \Delta v \rangle_{R1,R2}$. Conversely,
the most perturbing configurations are the ones with
lowest $q$ values, the equal--mass case ($q=0.5$)
leading to the lowest $\langle \Delta v \rangle_{R1,R2}$.
This result is logical when considering the structure of 
the secular term of the perturbing potential \citep[see for example
the Appendix of][]{mori}: the forced eccentricity term is proportional
to $(1-2q)$, i.e. it vanishes for the equal--mass case (where only
short period perturbation terms remain). An alternative and
simplified, but partially correct explanation is that the
equal mass case is the one for which the radial excursion of both
stars from the centre of mass is the more limited, i.e., they
both stay at 0.5\,AU from it, whereas for smaller $m_2/m_1$ values,
star number 2 moves further away from
the centre of mass and ``enters'' deeper into the circumbinary disk.
Within the range of parameters which have been explored, the
most perturbed case is thus logically the one with
mass ratio $q = 0.1$ and $e_b = 0.5$.
In any case, even for different size impacting objects, one robust result
is that the presence of gas drag always strongly reduces the 
relative velocity between planetesimals. 
This can be clearly seen in Table I, where we display,
for one given set of binary parameters (the same as in Fig.\ref{f1}),
relative velocity values for the case without gas drag,
both from our simulations and from Moriwaki and Nakagawa (2004), and for the 
case with gas drag for a target planetesimal of size $R_1=5$km.
We consider both the
case of equal-size planetesimals with $R_2=R_1 = 5$ km and 
different size planetesimals with $R_1 = 5$km and $R_2 = 2.5$ km.

As discussed in the introduction,
all $\langle \Delta v \rangle_{R1,R2}$ values have to be compared to two
critical values in order to determine the 
collisional evolution of a planetesimal swarm. One is the 
escape velocity of the target $v_{esc(R1)}$, which can 
easily be derived from the body's radius and density. Only if
$\langle \Delta v \rangle_{R1,R2} <<  v_{esc(R1)}$ can
runaway growth occur. The other crucial threshold value
is $v_{ero}$, the threshold velocity beyond which the net outcome
of an impact is mass erosion instead of accumulation into a
larger body. To estimate the value of $v_{ero(R1,R2)}$ for a given 
projectile and target is a complex task. It has to be derived 
from models of cratering and fragmentation which, at present, 
give uncertain results due to our poor knowledge of the 
size--strength scaling law.  
As discussed in \citet{the06}, there are many different 
models in the literature giving the value of the critical
specific shattering energy $Q*$ and the
size distribution of fragments 
as a function of the projectile and target size, of composition 
and of impact velocity. To be conservative, we consider as 
in \citet{the06} three different estimates of 
$v_{ero(R1,R2)}$ computed according to \citet{hol94}, \citet{mar95} and
\citet{benz99} prescriptions for $Q*$ and \citet{the03} prescription
for the fragmented and cratered masses.
Out of these we take the maximum value that 
gives an optimistic estimate of the region where accretion is 
possible, and the minimum and more conservative value. 

\begin{table*}
\begin{center}
\begin{tabular}{|r|c|c|c|c|}
\hline
\hline
r (AU)& No gas& No gas & Gas drag & Gas drag\\
      &(our simulations) &\citep{mori}  & R1=R2=5km& R1=5km R2=2.5km \\
\hline
4.25&849\,m.s$^{-1}$ &2040\,m.s$^{-1}$&18\,m.s$^{-1}$&12\,m.s$^{-1}$\\
7.75&283\,m.s$^{-1}$&754\,m.s$^{-1}$&3\,m.s$^{-1}$&9\,m.s$^{-1}$  \\
11.75&49\,m.s$^{-1}$&444\,m.s$^{-1}$&5\,m.s$^{-1}$&20\,m.s$^{-1}$ \\
\hline
\hline
\end{tabular}
\caption{Examples of impact velocity values $<\Delta v>$ obtained
with different models, for the same example binary configuration as
in Fig.\ref{f1} ($q=0.2$, $e_b=0.4$). 
}
\end{center}
\end{table*}

When we compare $\langle \Delta v \rangle_{R1,R2}$ to $v_{esc(R1)}$ and
$v_{ero(R1,R2)}$ we can outline 3 different accretional modes:
\begin{itemize}
\item
1) $\langle \Delta v \rangle <<  v_{esc(R1)}$. In this case,
runaway accretion can proceed as in the 'standard' unperturbed
scenario. The additional perturbations of the binary are not 
large enough to cancel out the 
gravitational focusing factor $(v_{esc}/\langle \Delta v \rangle)^2$ 
of the target, which is the source of this
fast growth mode e.g. \citep{lis93}. 
\item
2) For values of  $\langle \Delta v \rangle$ larger than 
$ v_{esc}$ but still smaller than $v_{ero}$,
planetary accretion is still possible but runaway growth will either 
not occur at all or it will start only when large bodies have formed 
such that $ v_{esc}>\langle \Delta v \rangle$.
In this case planetesimals will accumulate into planetary bodies 
but at a significantly
slower rate.
\item
3) If $\langle \Delta v \rangle$ exceeds
$v_{ero}$, the majority of collisions end up with cratering
and fragmentation that overcome accretion.  The planetesimal
population, instead of growing in size, will be slowly 
ground down to dust. 
\end{itemize}
 
For sake of clarity we display
our numerical predictions on planetesimal accretion in 4 
summarizing graphs. In each of them we display,
for all explored values of $q$ and $e_b$, 
the radial distance $r_l$ beyond which planetesimal
accretion is possible, for a population of objects having
an initial size $R_1=R_{min}$, all the way up to the maximal size
of 100km considered in our simulations.
Beyond that size, accretion dominates over erosion for
all explored cases.
In other words, $r_l$ is the limit beyond which
$\langle \Delta v \rangle_{R1,R2} < v_{ero(R1,R2)}$ for all 
($R_1,R_2$) pairs in the $R_1 > R_{min}$ range.
We consider two cases for the starting $R_{min}$ size:
one ``standard'' case with $R_{min}=5$km, and one
``small planetesimals'' case for which $R_{min}=0.5$km.
Note that the value $R_{min} = 5$ km is the commonly accepted 
minimum size for planetesimals: Independently of their formation 
process, when they reach this size they 
detach from the gas of the disk and move on independent 
Keplerian orbits, only perturbed by the gas.  This value is however approximate 
and this is why we also explore
a lower value of 0.5km for the initial planetesimal size.
We also consider the two maximum and minimum values derived
from the different prescriptions considered for $v_{ero(R1,R2)}$.

\begin{figure}
\includegraphics[width=8.5cm]{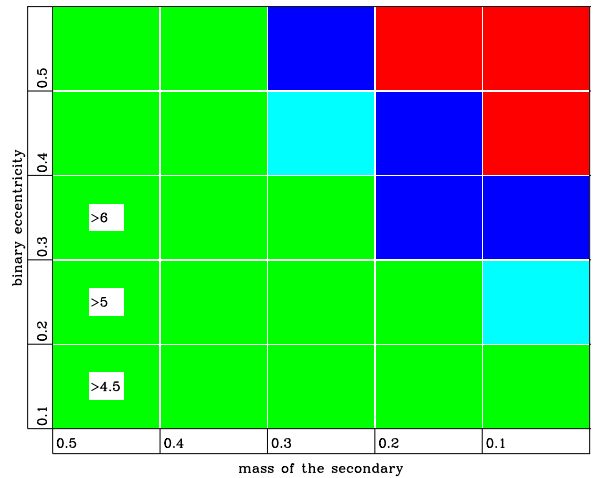}
\caption[]{Map of the radial distance $r_l$ beyond which planetary
formation is possible, i.e. where all ($R_1,R_2$) encounters lead
to mass accretion ($\langle \Delta v \rangle_{R1,R2} < v_{ero(R1,R2)}$)
for all objects of size $R_1> R_{min}=5$km, as a function of the binary mass ratio 
$q$ and binary eccentricity $e_b$. 
The color coding used in Figures 3-6 is the  following:
 
-{\it green}: $r_l\leq4$AU (the inner edge of our planetesimal disc)

-{\it pale blue}: 4AU$<r_l\leq 6$AU

-{\it dark blue}: 6AU$\leq r_l \leq 9$AU

-{\it red}: 9AU$\leq r_l<12$AU

-{\it black}: $r_l \geq 12$AU (the outer edge of our planetesimal disc)

The radial distance given at the center of some rectangles is 
the minimum value beyond which 
{\it runaway} accretion
is possible (i.e. $\langle \Delta v \rangle_{R1,R2}
<<  v_{esc(R1)}$) for all $R_1>5$km.
}
\label{highVero_bigplan}
\end{figure}

Fig.\ref{highVero_bigplan} displays the most favorable case
for accretion, where we assume $R_{min}=5$km and
adopt the highest, and thus less restrictive value for 
$v_{ero}$. It can be seen that, in spite of
the binary gravitational perturbations and of the 
periastra de--phasing of  
different size planetesimals, for mass ratios $q=$0.5 and 0.4
planet formation proceeds in all the 
regions of the disk and for all binary eccentricities.
The only region where accretion is inhibited is within the 
inner tidal gap of the disk.
For smaller values of $q$ and high binary eccentricities $e_b$,
the inner border for accretion shifts to larger radial 
distances. For $q = 0.1$ and $e_b = 0.5$ 
planet formation can occur only beyond 10 AU: strong secular perturbations
due to the large eccentricity and low mass ratio of the binary
prevent planetesimal accumulation closer to the barycenter. 
Note however that for most cases where we found accretion to
be possible the growth mode can probably not be runaway, or at least
not $100\%$ runaway. This is because $\langle \Delta v \rangle_{R1,R2}$
values are not $<<  v_{esc(R1)}$ for all explored $R_1$ values. The only
exceptions, where unperturbed runaway accretion is possible, are
obtained for the equal--mass case $q=0.5$ and
low binary eccentricities $e_b\leq 0.3$
(the white rectangles in Fig.\ref{highVero_bigplan}).

\begin{figure}
\includegraphics[width=8.5cm]{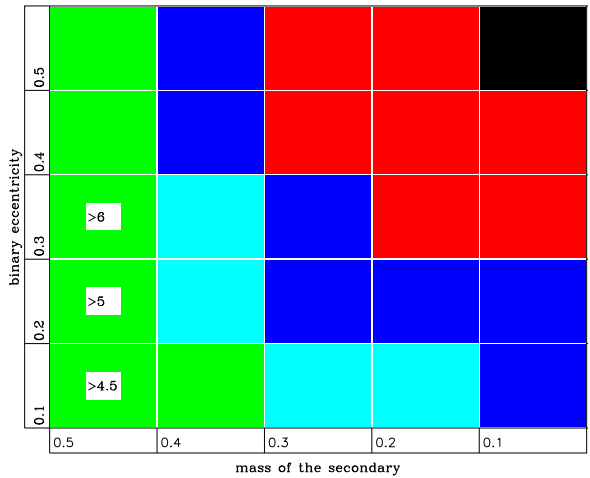}
\caption[]{Same as in  Fig.\ref{highVero_bigplan} but assuming
the minimum, and thus
more restrictive value of $v_{ero(R1,R2)}$.
}
\label{lowVero_bigplan}
\end{figure}

For a more conservative choice of the erosion velocity (Fig.\ref{lowVero_bigplan}), 
the inner limit for planetesimal accretion moves significantly outward,
as expected: only for $q = 0.5$, i.e. equal mass stars, 
is accretion possible at any radial distance,
independently of $e_b$.

\begin{figure}
\includegraphics[width=8.5cm]{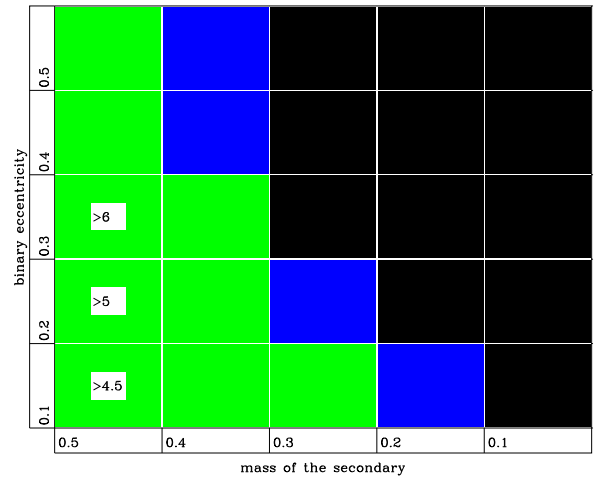}
\caption[]{Same as in Fig.\ref{highVero_bigplan} (``optimistic''
high $v_{ero}$ value) but assuming a 
minimum value of $R1$ equal to 0.5 km (small planetesimals).
}
\label{highVero_smallplan}
\end{figure}

\begin{figure}
\includegraphics[width=8.5cm]{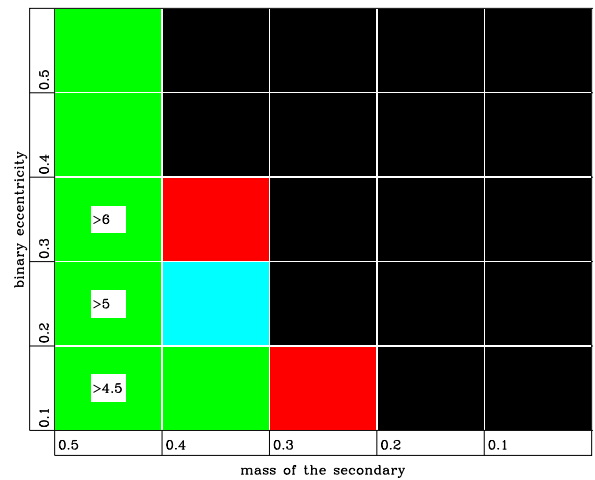}
\caption[]{Same as in  Fig.\ref{lowVero_bigplan} (restrictive
low $v_{ero}$ value), but assuming a 
minimum value of $R1$ equal to 0.5 km (small planetesimals).
}
\label{lowVero_smallplan}
\end{figure}

The cases with smaller 0.5km initial planetesimals appear to be
more critical (\ref{highVero_smallplan} and Fig.\ref{lowVero_smallplan}).
While for mass ratios
of 0.5 and 0.4 planet formation is (as in the $R_{min}=5$km case)
possible almost all over the disk, 
when $q$ becomes smaller than 0.4 in most cases accretion of small
planetesimals is inhibited within 12 AU from the stars. 
The main reason for this accretion inhibiting behaviour is that
for very small bodies (in the sub--kilometer range), orbital
phasing due to gas drag is very efficient and, as a consequence, 
the {\it differential}
phasing between objects of different sizes is very strong. This leads
to high $\langle \Delta v \rangle_{R1,R2}$, especially when compared to the very
low erosion threshold velocities corresponding to such small objects.

It is important to note that 
our results can be rescaled to different values of
the reference gas density $\rho_{\rm g0}$. Indeed,
the behaviour of a population of planetesimals of size $R$
in a gas disk of density  $\rho_{\rm g}$ is similar to the
behaviour of a population of planetesimals of size $R/X$
in a gas disk of density  $\rho_{\rm g}/X$ (see Equs.1\&2).
As an example, reducing by a factor 10 
the size of the bodies is equivalent to model the evolution of 
larger bodies but increase the gas density by a factor 10.
As a consequence, another way of reading
Figs.\ref{lowVero_smallplan} and \ref{highVero_smallplan} is that
in denser circumbinary disks, planetary formation 
may be easily inhibited even when starting from ``standard''
$R_{min}=5$km initial planetesimals.
Note however that
the rescaling is not fully straightforward, because
5km and 0.5km objects do not have the same threshold velocity
$v_{ero}$.

\section{Conclusions and Perspectives}

With a full numerical approach we have investigated planetesimal accretion and
the possibility of planet formation in circumbinary disks. The trajectories 
of 10000 planetesimals have been integrated over time.
The evolution of one crucial parameter, i.e. the distribution
of mutual encounter velocities within the system,
is numerically explored for a wide range of binary parameters,
specifically the mass ratio $q$ and orbital eccentricity $e_b$, the
semi-major axis being fixed at 1AU.
From these data we map the
minimum radial distance from the barycenter of the two stars beyond which 
accretion is always possible as a function of the binary mass ratio and 
eccentricity. Planets can form only in regions of the disk beyond this
distance. 

Our results significantly differ from those of previous studies
\citep{mori} mainly for two reasons:
\begin{itemize}

\item The use of the $dv\propto e\times v_{Kep}$
formula that relates the forced eccentricity of the binary companion to the 
average collisional velocity does not work properly when there is
a significant radial mixing of planetesimals. The comparison of our
full numerical results with those obtained from the forced eccentricity 
differ by almost a factor two. The estimate based on the 
forced eccentricity leads to artificially higher impact velocities. 

\item We include in our simulations the crucial effect of the
drag force due to the gaseous component of the disk. This
significantly affects the relative velocities between planetesimals: 
the $\langle \Delta v \rangle$ between similar-size bodies are significantly
reduced by the alignment of the periapses, but different-size planetesimals
are aligned towards different directions and they may experience 
higher impact velocities.

\end{itemize}

Summarizing all our results on equal and different-size planetesimals 
we outline the regions of the disk where accretion is always possible. 
We find that for a ``standard'' system where planetesimals have an
initial size of$\sim 5$km, accretion is always possible, in the considered
$>4$AU region, for equal-mass binaries. For lower $m_2/m_1$ mass ratios,
results significantly depend on 
the threshold velocity $v_{ero(R1,R2)}$ above which an impact
by an impactor $R_2$ on a target $R_1$ leads to net erosion instead of
net accretion. A general result is that lower $m_2/m_1$
and higher $e_b$ values result in higher $\langle \Delta v \rangle_{(R1,R2)}$
and are thus less favorable to planetesimal accretion.

One aspect of the problem that needs further investigation is the 
initial orbital distribution of planetesimals.
This parameter is linked to the difficult issue of
when and how they detach from the gas. 
While fully formed planetesimals may be marginally affected by the spiral 
waves of the disk caused by the binary, smaller bodies would tend to 
follow the gas streamlines when they are small enough to be coupled to 
the gas. This is a complex problem since it is strongly tied to the 
planetesimal formation process. How is the dust aggregation process 
influenced by the gas waves and by the stellar secular perturbations?
Is the gravitational instability still possible? These questions may 
be answered within a model capable to handle gas and dust at the same 
time.  
Such a complete model not being available yet, we chose to make
in our simulations the simplifying assumption that at $t=0$ planetesimal orbits
are circular, i.e., that their proper eccentricity is equal to the forced one.
However, we think that this choice may not have too dramatic consequences 
on the estimation of average relative velocities. Indeed,
we observe that the further evolution of
the system, under the secular perturbations of the stars and of gas drag ,
leads to a steady state where the proper eccentricity
is damped down and the periapses are aligned depending on body size.
This steady state only weakly depends on the initial orbital
parameters, especially for smaller planetesimals. 
In addition, 
since the major contribution to impact velocities is due to the 
different alignment of different-size planetesimals, rather than due to the 
absolute value of eccentricity, another choice of 
initial planetesimal orbits may not 
lead to significantly different results.

We considered
a binary semimajor axis $a_b$ of 1 AU in our simulations, but equation
3 for the gas-free case
can be easily scaled to larger (or smaller) values of $a_b$.
However, when we include  gas drag, such an easy scaling 
is no longer possible. The dynamical evolution of planetesimals
depends on the local gas density, which in turns depends on
the radial distance to the star. As a consequence, the evolution
of a planetesimal $R_1$ at a distance $r$ of the centre of mass
of a binary of separation $a_b$ is not equivalent to the evolution
of the same planetesimal at distance $r/2$ around a binary of separation
$a_b/2$ (even when corrected by a timescale correcting factor), because
the gas density is not the same at these 2 locations. Furthermore,
there is no reason to assume that the global gas disk profile is the same
around binaries with different orbital separations. As an example,
if for $a_b = 1$ AU it may be reasonable to 
adopt the minimum mass solar nebula density $\rho_0$ at 4 AU, in 
a disk around a binary with
semimajor axis of $a_b = 5$ AU, adopting the same 
density $\rho_0$ at 20 AU appears inconsistent.
In other words, while the relative velocity in absence of 
gas scales with $a_b$ because of its dependence on the Keplerian velocity, 
its density does not necessarily scale 
with $r$ when gas drag is considered. 
As a consequence, our results may be applied to disks around 
binaries with small separations, but they cannot be scaled to large 
$a_b$. 

An additional effect that we neglect in our simulations is 
the physical outcome of mutual collisions (be it accretion
or erosion).  As noted in the introduction,
a model capable to handle at the same time the collisional and 
dynamical evolution of a swarm of planetesimals
is still far from the present computing capabilities. 

Our work must be considered a step forward of a better understanding of 
planet formation in circumbinary disks. There are certainly
aspects of the problems that need further investigation in future papers 
like planetary accretion in habitable regions around binaries,
planetary migration driven by disk interaction and by planetesimal encounters.

\end{document}